\documentclass[pre,amssymb,amsmath,twocolumn,aps,showpacs,superscriptaddress]{revtex4}
\usepackage{amsmath,graphicx}

\usepackage[dvips]{epsfig}
\usepackage[usenames,dvipsnames]{color}
\usepackage[caption=false]{subfig}
\usepackage{verbatim}
\newcommand{\nmax}{n_{\mathrm{max}}} 
\newcommand{\peq}{p^{\mathrm{eq}}_n(\mu)}
\newcommand{\nav}{\langle n \rangle}
\newcommand{\nvar}{\langle n^2 \rangle - \langle n \rangle^2}

\begin{document}
\title{Comment on ``Generalized exclusion processes: Transport coefficients''}
\author{T. Becker}
\email{thijsbecker@gmail.com}
\affiliation{Hasselt University, B-3590 Diepenbeek, Belgium}
\author{K. Nelissen}
\affiliation{Departement Fysica, Universiteit Antwerpen, Groenenborgerlaan 171, B-2020 Antwerpen, Belgium}
\affiliation{Hasselt University, B-3590 Diepenbeek, Belgium}
\author{B. Cleuren}
\affiliation{Hasselt University, B-3590 Diepenbeek, Belgium}
\author{B. Partoens}
\affiliation{Departement Fysica, Universiteit Antwerpen, Groenenborgerlaan 171, B-2020 Antwerpen, Belgium}
\author{C. Van den Broeck}
\affiliation{Hasselt University, B-3590 Diepenbeek, Belgium}
\date{\today}
\begin{abstract}
In a recent paper Arita \textit{et al.}~[Phys.~Rev.~E~\textbf{90}, 052108 (2014)] consider the transport properties of a class of generalized exclusion processes. Analytical expressions for the transport-diffusion coefficient are derived by ignoring correlations. It is claimed that these expressions become exact in the hydrodynamic limit. In this Comment, we point out that (i) the influence of correlations upon the diffusion does not vanish in the hydrodynamic limit, and (ii) the expressions for the self- and transport diffusion derived by Arita \textit{et al.}~are special cases of results derived in [Phys.~Rev.~Lett.~\textbf{111}, 110601 (2013)].
\end{abstract}
\maketitle

\begin{figure}
\centering
\includegraphics[width=\columnwidth]{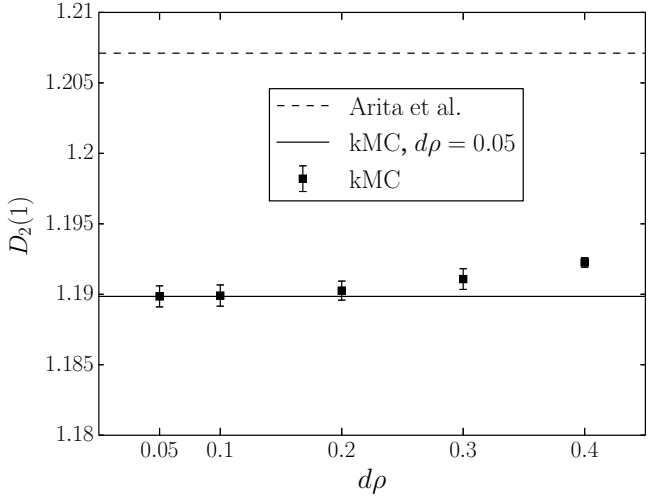}
\caption{Transport diffusion $D_2$ in one dimension as a function of $d \rho$, for $L_x=20$ and $\rho = 1$.}
\label{fig::convdc}
\end{figure}

\begin{figure}
\centering
\includegraphics[width=\columnwidth]{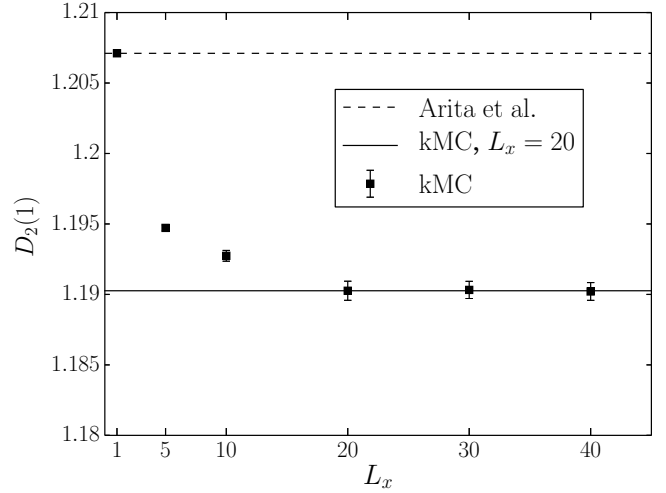}
\caption{Transport diffusion $D_2$ in one dimension as a function of $L_x$, for $\rho = 1$ and $d \rho = 0.1$. Value at $L_x=1$ is calculated analytically.}
\label{fig::convL}
\end{figure}

\begin{figure}
\centering
\includegraphics[width=\columnwidth]{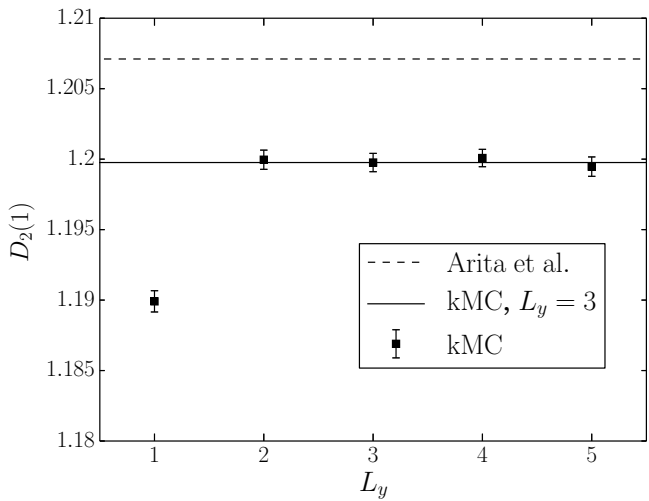}
\caption{Transport diffusion $D_2$ in two dimensions as a function of $L_y$, for $d \rho = 0.1$, $\rho = 1$, and $L_x = 20$. Periodic boundary conditions are imposed in the $y$ direction.}
\label{fig::convLy}
\end{figure}

\begin{figure}
\centering
\includegraphics[width=\columnwidth]{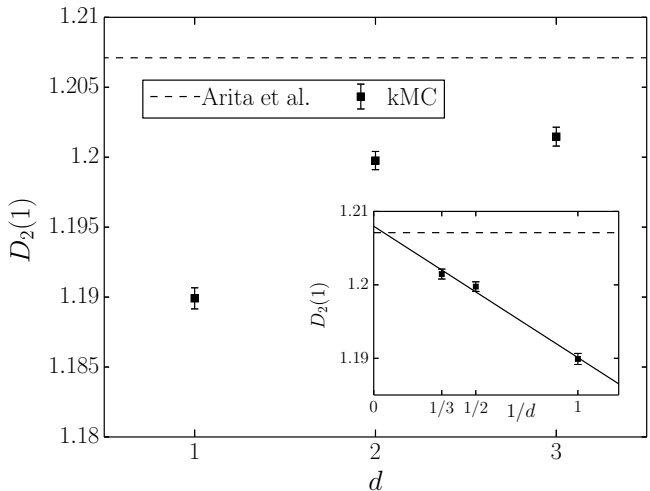}
\caption{Transport diffusion $D_2$ in one, two, and three dimensions, for $\rho = 1$, $d \rho = 0.1$, $L_x = 20$, and $L_y = L_z = 3$. Periodic boundary conditions are imposed in the $y$ and $z$ directions. The inset shows the same data plotted as a function of $1/d$, together with a $1/d$ fit (obtained with Mathematica).}
\label{fig::Dt_dim}
\end{figure}

\begin{figure}
\centering
\includegraphics[width=\columnwidth]{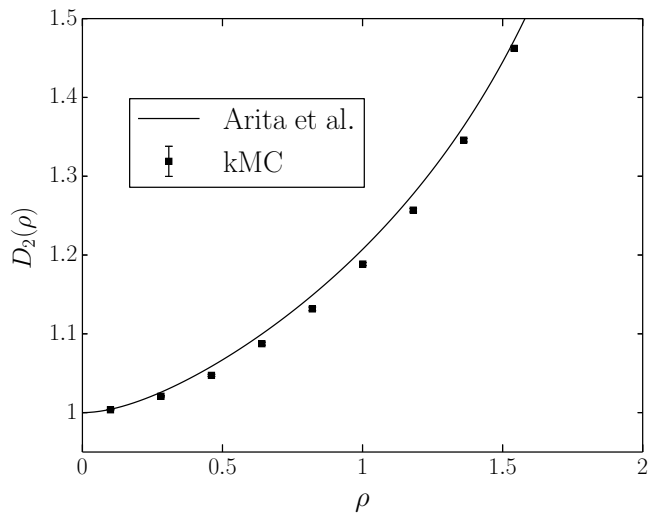}
\caption{Transport diffusion $D_2(\rho)$ in one dimension for $L_x = 50$ and $d \rho = 0.18$.}
\label{fig::Dt_rho}
\end{figure}

In a recent paper \cite{PRE_Arita}, Arita \textit{et al.}~derived analytical expressions for the transport-diffusion coefficient in a generalized exclusion process. Their derivation depends crucially on the assumption that correlations in the dynamics can be ignored up to first order in the concentration gradient, at least in the hydrodynamic regime. Since this is the regime in which the transport diffusion is defined, it was claimed that these expressions are exact and valid in all spatial dimensions. In this Comment we show numerically that correlations do, in fact, influence the transport diffusion. As a result the quoted expressions are not exact and the transport diffusion depends on the dimension. Furthermore, the  expressions for both the self- and transport diffusion derived by Arita \textit{et al.}~are special cases of results derived in Ref.~\cite{PRL2013}.

The generalized exclusion process is defined as follows. Consider a $d$-dimensional cubic lattice in which every lattice site can contain between zero and $k$ particles. Each particle attempts to hop to one of its neighbor sites with unit rate. The hopping attempt is successful when the target site is occupied by less than $k$ particles. In the hydrodynamic regime, i.e., on large length and time scales and for small enough gradients  \cite{ADVPHYSalanissila2002}, the particle flux $j$  through the system in response to a concentration gradient $d \rho / dx$ is expected to obey Fick's first law:
 \begin{equation}\label{eq::DtFick}
j = - D_k(\rho) \frac{d \rho}{d x}.
\end{equation}
The transport-diffusion coefficient $D_k(\rho)$ will depend on the maximum occupancy $k$ and the particle density $\rho$.  
In the following, we will focus on the ``simplest" extension of the exclusion process, allowing up to $2$ particles per site,  $k=2$. 
When neglecting spatial correlations Arita \textit{et al.}~obtained the following result: 
\begin{equation}\label{eq::result_arita}
D_2(\rho) = \frac{1+ \rho+\sqrt{1+ 2 \rho - \rho^2} }{2 \sqrt{1+2\rho-\rho^2}}.
\end{equation} 

Our numerical evaluation of the transport diffusion is based on a direct simulation of the dynamics by kinetic Monte Carlo (kMC), cf.~Refs.~\cite{PRL2013,PREbecker} for a description of the simulation methods. In two and three dimensions the algorithm of Schulze \cite{schulze2002kinetic} is used. The system is connected on its left and right boundary to particle reservoirs at, respectively, densities $\rho + d \rho/2$ and $\rho - d \rho/2$. $D_k(\rho)$ is obtained via Eq.~\eqref{eq::DtFick}  from the observed  particle flux through the system.  

We first verify convergence of our simulation results to the  linear response (small gradient) and hydrodynamic  (large system size) regime.  The  distance between two neighboring sites is set equal to one. We focus on the case with density $\rho = 1$. For a one-dimensional lattice consisting of $L_x$ sites, we evaluate the diffusion coefficient $D_2(1)$   as a function of  $d \rho$ for $L_x=20$, cf. Figure \ref{fig::convdc}. One-sigma error bars are shown in all figures.
Convergence to the linear response regime is found from $d \rho \approx 0.1$. In the same way, we identify the hydrodynamic regime by evaluating $D_2(1)$ for different system lengths $L_x$, cf. Figure \ref{fig::convL}. There is convergence from $L_x \approx 20$. The value at $L_x = 1$ is calculated analytically \cite{PRL2013}. Note that it is equal to the uncorrelated result Eq.~\eqref{eq::result_arita}. In the two-dimensional system we impose periodic boundary conditions in the $y$ direction. As is illustrated in Figure \ref{fig::convLy}  convergence to the hydrodynamic limit already sets in around $L_y \approx 3$. We proceed in a similar way for a three dimensional system.

We now discuss our main results. In Figure \ref{fig::Dt_dim} we plot the thus estimated transport diffusion $D_2(1)$ in one, two, and three dimensions. $D_2(1)$ is clearly different from the result of Ref.~\cite{PRE_Arita}, and depends on the dimension. One can argue that in the limit of infinite dimension correlations have no influence. In the inset of Figure \ref{fig::Dt_dim} we plot a $1/d$ fit to the numerical data. The extrapolation to the $d \rightarrow \infty$ limit is close to the uncorrelated result. The same $1/d$ scaling was found in \cite{becker2015current}.
As a further illustration, we plot $D_2(\rho)$ in one dimension in Figure \ref{fig::Dt_rho} together with  Eq.~\eqref{eq::result_arita}. Even though the $\rho$-dependence is qualitatively reproduced, the numerically obtained value of the diffusion coefficient is lower. From extensive numerical simulations \cite{PRL2013,PREbecker,becker2015current} we suspect that correlations always lower the diffusion. We refer to \cite{PREbecker} for a detailed discussion of the influence of correlations upon the diffusion. The exact analytical result for $D_2(\rho)$ in one dimension for $L_x = 2$ is discussed in \cite{EPJSTbecker}. Correlations are already present for $L_x = 2$ (in contrast to $L_x = 1$). One finds analytically that the uncorrelated result is higher than the exact result for all concentrations. The same qualitative behavior as in the $L_x \rightarrow \infty$ limit, Figure \ref{fig::Dt_rho}, is observed.

Finally, we note that the analytical expressions for the self- and transport diffusion obtained in Ref.~\cite{PRE_Arita} are special cases of results presented in Ref.~\cite{PRL2013}. The model from \cite{PRL2013} is a one-dimensional lattice gas, where each site can contain a maximum of $n_{\mathrm{max}}$ \emph{interacting} particles. A site containing $n$ particles has equilibrium free energy $F(n)$. This free energy includes contributions due to the interaction between different particles. If the system is in equilibrium at chemical potential $\mu$, the probability to find $n$ particles on a lattice site is
\begin{equation}\label{eq::probeq}
p^{\mathrm{eq}}_n(\mu) = \left[ \mathcal{Z}(\mu)\right]^{-1} e^{- \beta \left[F(n)-\mu n\right]},
\end{equation}
with $\beta=(k_B T)^{-1}$, $k_B$ the Boltzmann constant, $T$ the temperature, and \( {\mathcal{Z}} \) the grand canonical partition function:
\begin{equation}
{\mathcal{Z}}(\mu) = \sum_{n=0}^{\nmax} e^{- \beta \left[F(n)-\mu n\right]}.
\end{equation}
Averages over the equilibrium distribution Eq.~\eqref{eq::probeq} are denoted by $\langle \cdot \rangle$, e.g., $\nav (\mu) = \sum_{n=0}^{\nmax} n \peq$ is the average number of particles on a lattice site. Particles jump from a site containing $n$ particles to a site containing $m$ particles with rate $k_{nm}$. This rate should obey local detailed balance w.r.t.~$\peq$.
If one ignores all correlations, the self-diffusion $D_s(\mu)$ and the transport diffusion $D_t(\mu)$ are found to be \cite{PRL2013}:
\begin{equation}
D_s(\mu) = \frac{\langle k \rangle}{\nav}, \quad D_t(\mu) = \frac{\langle k \rangle}{\nvar}.
\end{equation}
The results of \cite{PRE_Arita} are recovered by setting $\nmax \equiv k$ and
\begin{equation}\label{rates}
k_{nm} =
\left\{ \begin{array}{ll}
n & \mathrm{if}\;\; m < \nmax \\
0 & \mathrm{if}\;\; m \geq \nmax
\end{array} \right.
\end{equation}
and by specifying that the particles are noninteracting. In this case, the free energy reduces to $F(n) = \beta^{-1} \ln(n!) + c n$ with $c$ a temperature dependent constant. The equilibrium distribution becomes:
\begin{equation}
\peq = \frac{e^{-\beta (c-\mu) n}}{n! {\mathcal{Z}}(\mu)},
\end{equation}
which is the result from \cite{PRE_Arita} with $\lambda = e^{-\beta (c-\mu) }$ (see also \cite{kipnis,SchutzJPA,basuPRE}). 
The average jump rate is then:
\begin{equation}
\langle k \rangle =  \sum_n \sum_m k_{nm} p^{\mathrm{eq}}_{n} p^{\mathrm{eq}}_{m} = \navÊ[1 - p^{\mathrm{eq}}_{\nmax}].
\end{equation}
Hence $D_s = 1 - p^{\mathrm{eq}}_{\nmax}$. This is the result from Ref.~\cite{PRE_Arita} for the self-diffusion, parametrized as a function of $\mu$  instead of $\lambda = e^{-\beta (c-\mu) }$. A similar, but more involved, calculation can be performed to show the equivalence of the expressions for the transport diffusion.  

To conclude, there is no guarantee that neglecting correlations leads to an exact result for the transport diffusion for generalized exclusion processes; see also the discussion in Ref.~\cite{PREbecker}. 
Uncorrelated expressions for the self- and transport diffusion are given in Ref.~\cite{PRL2013}, of which the results in \cite{PRE_Arita} are a special case.

\begin{acknowledgments}
This work was supported by the Flemish Science Foundation (Fonds Wetenschappelijk Onderzoek), Project No.~G.0388.11. The computational resources and services used in this work were provided by the VSC (Flemish Supercomputer Center), funded by the Hercules Foundation and the Flemish Government, department EWI.
\end{acknowledgments}


\begin{thebibliography}{10}
\expandafter\ifx\csname natexlab\endcsname\relax\def\natexlab#1{#1}\fi
\expandafter\ifx\csname bibnamefont\endcsname\relax
  \def\bibnamefont#1{#1}\fi
\expandafter\ifx\csname bibfnamefont\endcsname\relax
  \def\bibfnamefont#1{#1}\fi
\expandafter\ifx\csname citenamefont\endcsname\relax
  \def\citenamefont#1{#1}\fi
\expandafter\ifx\csname url\endcsname\relax
  \def\url#1{\texttt{#1}}\fi
\expandafter\ifx\csname urlprefix\endcsname\relax\def\urlprefix{URL }\fi
\providecommand{\bibinfo}[2]{#2}
\providecommand{\eprint}[2][]{\url{#2}}

\bibitem[{\citenamefont{Arita et~al.}(2014)\citenamefont{Arita, Krapivsky, and
  Mallick}}]{PRE_Arita}
\bibinfo{author}{\bibfnamefont{C.}~\bibnamefont{Arita}},
  \bibinfo{author}{\bibfnamefont{P.~L.} \bibnamefont{Krapivsky}},
  \bibnamefont{and} \bibinfo{author}{\bibfnamefont{K.}~\bibnamefont{Mallick}},
  \bibinfo{journal}{Phys. Rev. E} \textbf{\bibinfo{volume}{90}},
  \bibinfo{pages}{052108} (\bibinfo{year}{2014}).

\bibitem[{\citenamefont{Becker et~al.}(2013)\citenamefont{Becker, Nelissen,
  Cleuren, Partoens, and Van~den Broeck}}]{PRL2013}
\bibinfo{author}{\bibfnamefont{T.}~\bibnamefont{Becker}},
  \bibinfo{author}{\bibfnamefont{K.}~\bibnamefont{Nelissen}},
  \bibinfo{author}{\bibfnamefont{B.}~\bibnamefont{Cleuren}},
  \bibinfo{author}{\bibfnamefont{B.}~\bibnamefont{Partoens}}, \bibnamefont{and}
  \bibinfo{author}{\bibfnamefont{C.}~\bibnamefont{Van~den Broeck}},
  \bibinfo{journal}{Phys. Rev. Lett.} \textbf{\bibinfo{volume}{111}},
  \bibinfo{pages}{110601} (\bibinfo{year}{2013}).

\bibitem[{\citenamefont{Ala-Nissila et~al.}(2002)\citenamefont{Ala-Nissila,
  Ferrando, and Ying}}]{ADVPHYSalanissila2002}
\bibinfo{author}{\bibfnamefont{T.}~\bibnamefont{Ala-Nissila}},
  \bibinfo{author}{\bibfnamefont{R.}~\bibnamefont{Ferrando}}, \bibnamefont{and}
  \bibinfo{author}{\bibfnamefont{S.~C.} \bibnamefont{Ying}},
  \bibinfo{journal}{Adv. Phys.} \textbf{\bibinfo{volume}{51}},
  \bibinfo{pages}{949} (\bibinfo{year}{2002}).

\bibitem[{\citenamefont{Becker et~al.}(2014{\natexlab{a}})\citenamefont{Becker,
  Nelissen, Cleuren, Partoens, and Van~den Broeck}}]{PREbecker}
\bibinfo{author}{\bibfnamefont{T.}~\bibnamefont{Becker}},
  \bibinfo{author}{\bibfnamefont{K.}~\bibnamefont{Nelissen}},
  \bibinfo{author}{\bibfnamefont{B.}~\bibnamefont{Cleuren}},
  \bibinfo{author}{\bibfnamefont{B.}~\bibnamefont{Partoens}}, \bibnamefont{and}
  \bibinfo{author}{\bibfnamefont{C.}~\bibnamefont{Van~den Broeck}},
  \bibinfo{journal}{Phys. Rev. E} \textbf{\bibinfo{volume}{90}},
  \bibinfo{pages}{052139} (\bibinfo{year}{2014}{\natexlab{a}}).

\bibitem[{\citenamefont{Schulze}(2002)}]{schulze2002kinetic}
\bibinfo{author}{\bibfnamefont{T.~P.} \bibnamefont{Schulze}},
  \bibinfo{journal}{Phys. Rev. E} \textbf{\bibinfo{volume}{65}},
  \bibinfo{pages}{036704} (\bibinfo{year}{2002}).

\bibitem[{\citenamefont{Becker et~al.}(2015)\citenamefont{Becker, Nelissen, and
  Cleuren}}]{becker2015current}
\bibinfo{author}{\bibfnamefont{T.}~\bibnamefont{Becker}},
  \bibinfo{author}{\bibfnamefont{K.}~\bibnamefont{Nelissen}}, \bibnamefont{and}
  \bibinfo{author}{\bibfnamefont{B.}~\bibnamefont{Cleuren}},
  \bibinfo{journal}{New J. Phys.} \textbf{\bibinfo{volume}{17}},
  \bibinfo{pages}{055023} (\bibinfo{year}{2015}).

\bibitem[{\citenamefont{Becker et~al.}(2014{\natexlab{b}})\citenamefont{Becker,
  Nelissen, Cleuren, Partoens, and Van~den Broeck}}]{EPJSTbecker}
\bibinfo{author}{\bibfnamefont{T.}~\bibnamefont{Becker}},
  \bibinfo{author}{\bibfnamefont{K.}~\bibnamefont{Nelissen}},
  \bibinfo{author}{\bibfnamefont{B.}~\bibnamefont{Cleuren}},
  \bibinfo{author}{\bibfnamefont{B.}~\bibnamefont{Partoens}}, \bibnamefont{and}
  \bibinfo{author}{\bibfnamefont{C.}~\bibnamefont{Van~den Broeck}},
  \bibinfo{journal}{Eur. Phys. J. Special Topics}
  \textbf{\bibinfo{volume}{223}}, \bibinfo{pages}{3243}
  (\bibinfo{year}{2014}{\natexlab{b}}).

\bibitem[{\citenamefont{Kipnis and Landim}(1999)}]{kipnis}
\bibinfo{author}{\bibfnamefont{C.}~\bibnamefont{Kipnis}} \bibnamefont{and}
  \bibinfo{author}{\bibfnamefont{C.}~\bibnamefont{Landim}},
  \emph{\bibinfo{title}{Scaling Limits of Interacting Particle Systems}}
  (\bibinfo{publisher}{Springer, New York}, \bibinfo{year}{1999}).

\bibitem[{\citenamefont{Sch{\"u}tz et~al.}(1996)\citenamefont{Sch{\"u}tz,
  Ramaswamy, and Barma}}]{SchutzJPA}
\bibinfo{author}{\bibfnamefont{G.~M.} \bibnamefont{Sch{\"u}tz}},
  \bibinfo{author}{\bibfnamefont{R.}~\bibnamefont{Ramaswamy}},
  \bibnamefont{and} \bibinfo{author}{\bibfnamefont{M.}~\bibnamefont{Barma}},
  \bibinfo{journal}{J. Phys. A: Math. Gen.} \textbf{\bibinfo{volume}{29}},
  \bibinfo{pages}{837} (\bibinfo{year}{1996}).

\bibitem[{\citenamefont{Basu and Mohanty}(2010)}]{basuPRE}
\bibinfo{author}{\bibfnamefont{U.}~\bibnamefont{Basu}} \bibnamefont{and}
  \bibinfo{author}{\bibfnamefont{P.~K.} \bibnamefont{Mohanty}},
  \bibinfo{journal}{Phys. Rev. E} \textbf{\bibinfo{volume}{82}},
  \bibinfo{pages}{041117} (\bibinfo{year}{2010}).

\end{thebibliography}

\end{document}